\documentclass[english]{article}
\usepackage[latin9]{inputenc}
\usepackage{textcomp}
\usepackage{graphicx}
\usepackage{setspace}
\usepackage[numbers]{natbib}
\makeatletter
\newcommand{\lyxaddress}[1]{
\par {\raggedright #1
\vspace{1.4em} 
\noindent\par}
}

\usepackage{hyperref}

\makeatother

\usepackage{babel}
\begin{document}

\title{Formation of recurring slope lineae on Mars by rarefied gas-triggered
granular flows}

\author{Fr{\'e}d{\'e}ric Schmidt$^{1*}$, Fran{\c c}ois Andrieu$^{1}$, Fran{\c c}ois Costard$^{1}$,
\\
Miroslav Kocifaj$^{2,3}$, Alina G. Meresescu$^{1}$}
\maketitle

\lyxaddress{$^{1}$ GEOPS, Univ. Paris-Sud, CNRS, Universit{\'e} Paris-Saclay, Rue
du Belv{\'e}d{\`e}re, B{\^a}t. 504-509, 91405 Orsay, France (frederic.schmidt@u-psud.fr),
$^{2}$ ICA, Slovak Academy of Sciences, Dubravska Road 9, 845 03
Bratislava, Slovak Republic, $^{3}$Faculty of Mathematics, Physics,
and Informatics, Comenius University, Mlynska dolina, 842 48 Bratislava,
Slovak Republic}

Paper submitted to Nature Geoscience
\begin{abstract}
Active dark flows known as recurring slope lineae have been observed
on the warmest slopes of equatorial Mars. The morphology, composition
and seasonality of the lineae suggest a role of liquid water in their
formation. However, internal and atmospheric sources of water appear
to be insufficient to sustain the observed slope activity. Experimental
evidence suggests that under the low atmospheric pressure at the surface
of Mars, gas can flow upwards through porous Martian soil due to thermal
creep under surface regions heated by the sun and disturb small particles.
Here we present numerical simulations to demonstrate that such a dry
process involving the pumping of rarefied gas in the Martian soil
due to temperature contrasts can explain the formation of the recurring
slope lineae. In our simulations, solar irradiation followed by shadow
significantly reduces the angle of repose due to the resulting temporary
temperature gradients over shaded terrain and leads to flow at intermediate
slope angles. The simulated flow locations are consistent with observed
recurring slope lineae that initiate in rough and bouldered terrain
with local shadows over the soil. We suggest that this dry avalanche
process can explain the formation of the recurring slope lineae on
Mars without requiring liquid water or CO$_{2}$ frost activity.
\end{abstract}

Dark flow type RSL occurs in the warmest areas of Mars, during the
warm season \citep{McEwen_SeasonalFlowsWarm_Science2011,McEwen_Recurringslopelineae_NG2014}.
The activity of RSL seems to be linked with solar irradiance. In the
$\sim$11\textdegree S latitude of Melas Chasma, the activity is observed
mainly around Ls = 90\textdegree{} (Southern hemisphere Winter solstice)
at a slope oriented toward North and around Ls = 270\textdegree{}
(Southern hemisphere Summer solstice) at a slope oriented toward South
\citep{McEwen_Recurringslopelineae_NG2014}. This slope dependent
seasonal activity is also confirmed by more recent observations, including
Garni crater, central Melas chasma \citep{Chojnacki_Geologiccontextrecurring_JoGRP2016}.
Thus high local temperature or local irradiance seem to be the trigger
of the flow.

The temperature and insolation dependence has been recognized very
early, and has been mainly interpreted as an effect of humidity \citep{McEwen_SeasonalFlowsWarm_Science2011,McEwen_Recurringslopelineae_NG2014}.
Indeed, RSL activity occurs in the closest point to liquid water in
the phase diagram, at present time on Mars \citep{Chojnacki_Geologiccontextrecurring_JoGRP2016}.
Various sources of liquid water have been proposed : subsurface aquifers,
melting of ice dams, deliquescence of salt recharged by atmospheric
water \citep{Ojha_Spectralevidencehydrated_NG2015}. The transport
phenomenon itself seems to be very puzzling. Recently, metastable
boiling water \citep{Masse_Transportprocessesinduced_NG2016} has
been proposed. If liquid water, even transient, is currently present
on Mars, its habitability would be relatively high. 

\paragraph*{Wet versus dry formation}

Nevertheless, the location of RSL is near the equator which is the
driest place on Mars and where atmospheric water vapor is at the lowest
\citep{Smith_AnnualWaterVaporTES_JGR2002}. In this place, surface
condensation of atmospheric water vapor never occurs \citep{Vincendon_WaterIceMidLatitudeMars_JGR2010}.
Subsurface ice is not stable \citep{Schorghofer_DynamicIceAge_Nature2007},
as also confirmed by indirect detection \citep{Vincendon_TropicalIce_GRL2010}.
The source of water seems still a mystery because an internal source,
such as a subsurface aquifer, has also been excluded especially when RSL occurs near the crater
rim \citep{Chojnacki_Geologiccontextrecurring_JoGRP2016}. The actual
amount of atmospheric water required to recharge the RSL's source
each year seems not sufficient. The precise thermal calculations from
remote sensing thermal infrared measurement of the THEMIS instrument
show no evidence of liquid water \citep{Edwards_WaterContentRecurring_GRL2016}
and there is no spectroscopic direct evidence of liquid water \citep{Ojha_Spectralevidencehydrated_NG2015}.
Recently discovered diurnal CO$_{2}$ frost could also trigger the
flow \citep{Piqueux_Discoverywidespreadlow_JoGRP2016} by sublimation
in early morning. Nevertheless this process is not consistent with
RSL growths observed in Valles Marineris at Ls = 270\textdegree .
At this period, diurnal CO$_{2}$ frost is only present in the bright
terrains between Arsia Mons and Mangala Fosse \citep{Piqueux_Discoverywidespreadlow_JoGRP2016}.
We propose here a re-interpretation of the RSL features and a new
process to explain the RSL activity without involving phase change
of chemical compounds. This process aims to reconcile all available
data.

\begin{figure}
\includegraphics[width=1\textwidth]{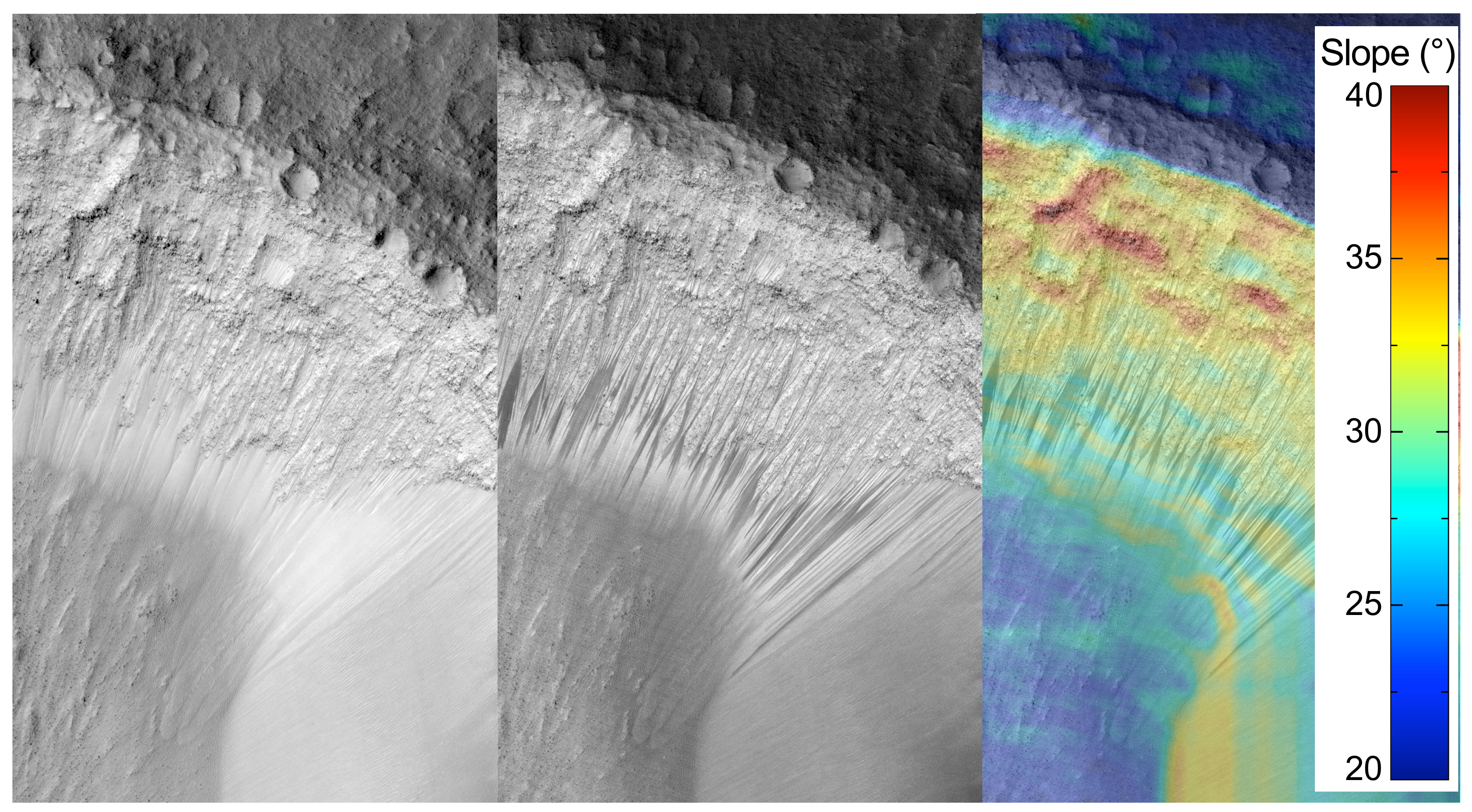}\caption{Evolution of RSL at Garni crater, Valles Marineris, Mars. Left: ESP\_029213\_1685
at Ls = 191.5\textdegree{} Middle: ESP\_031059\_1685 at Ls = 280.8\textdegree{}
Right: slope angle map using stereoscopic DTM (see sup. mat. for details)
Image MRO, HiRISE, NASA/JPL/University of Arizona. Scale: width=500
m. \label{fig:RSL-image}}
\end{figure}

Geomorphologically, the RSL consists of the following parts: (1) a
concentrated source from bedrock outcrops, (2) a linear channel that
is a few meters large and several 100 m long, (3) a terminal part with
elongated tongues (figure \ref{fig:RSL-image}). The main channel
exhibits a rather streak-like shape without levees (at the HiRISE
resolution of 25 cm). The source has been often attributed to the
outcrops but the outcrops themselves are not changing. Moreover, the
steepest parts of the walls always stay unchanged. Dark features appear
to be surprisingly young and do not exhibit any meandering, anastomosing
patterns or ramifications. Of particular interest, some sinuosities
are reported on some dry granular flows on Earth \citep{Shinbrot_DryFlow_PNAS2004}
but without any cyclic variations like meandering. These sinuosities
on RSL were previously described as a diversion around obstacles like
in some dry granular flows \citep{Mangold_DryWetDebrisFlow_Conf2008}.
The terminal point of RSL is usually determined by a slope angle of
\textasciitilde{}28\textdegree . In summary, RSLs exhibit a narrow
streak shape, an observed runout distance and distal morphology with
exactly the same corresponding morphological characteristics of most
terrestrial dry rock avalanches. Some laboratory simulations show
the possibility to have some levee-channel deposits on dry granular
flows \citep{Felix_DryGranularFlow_EPSL2004}. Unfortunately, the
resolution of HIRISE images is not enough to detect the morphometry
of levees at the scale of these RSL. A fluvial or viscous flow like
debris flows would show visible terminal levees at the HiRISE resolution
and a much larger channel \citep{Costard_GulliesDry_MarsConf2007}.
The lack of any lateral or terminal levees strongly favors a dry granular
flow hypothesis. RSLs appear darker than the surrounding terrains
for a few weeks and then they fade away. Vertical inverse grading
is a common process in clastic flows due to kinetic sorting \citep{Legros_CanDispersivePressure_JoSR2002,Armanini_Granularflowsdriven_JoHR2013,Kokelaar_Finegrainedlinings_EaPSL2014}.
We interpret the relative darker albedo appearance of the RSL as an
effect of the sorting of the grains during the flow. Finer grains could
also be ejected during the flow. The fading is then due to the finer
eolian grain deposition from the atmosphere. In conclusion, most of
the RSL exhibit morphological characteristics that do not necessarily
involve liquid phase for their formation, but are mostly in agreement
with dry granular flows like the ones typical of terrestrial rock
avalanches.

\begin{figure}
\includegraphics[width=1\textwidth]{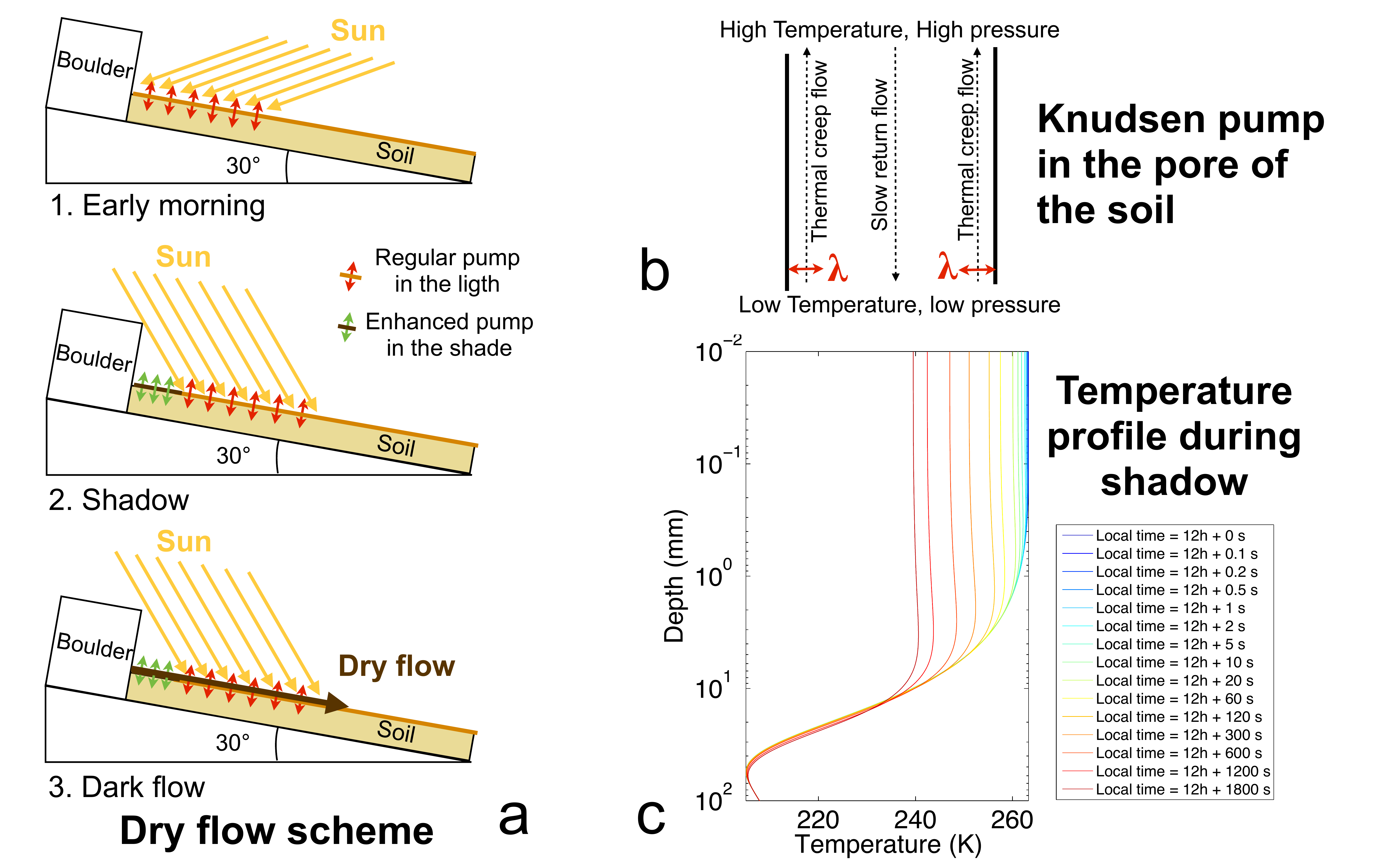}\caption{Knudsen pump in the Martian soil. 
(a) Scheme of the dry flow triggered by the Knudsen pump. The regular pump occurs in the irradiated soil.
The enhanced pump occurs few minutes after the appearance of the shadow
of a boulder. 
(b) Scheme of the thermal creep flow in the porous space
of the Martian soil, driven from the cold end to the hot end of the
tube, close to the wall. $\lambda$ is the free mean path of the gas, 
about 20 microns in the Martian soil. 
(c) Temperature profile at local 12h (before shadow) and after a shadowing
period (from 0.1 to 1800s after noon), for a facet at slope 30\textdegree{}
toward North at Ls = 90\textdegree . \label{fig:Scheme-Dry-Flow}}
\end{figure}

\begin{figure}
\includegraphics[width=1\textwidth]{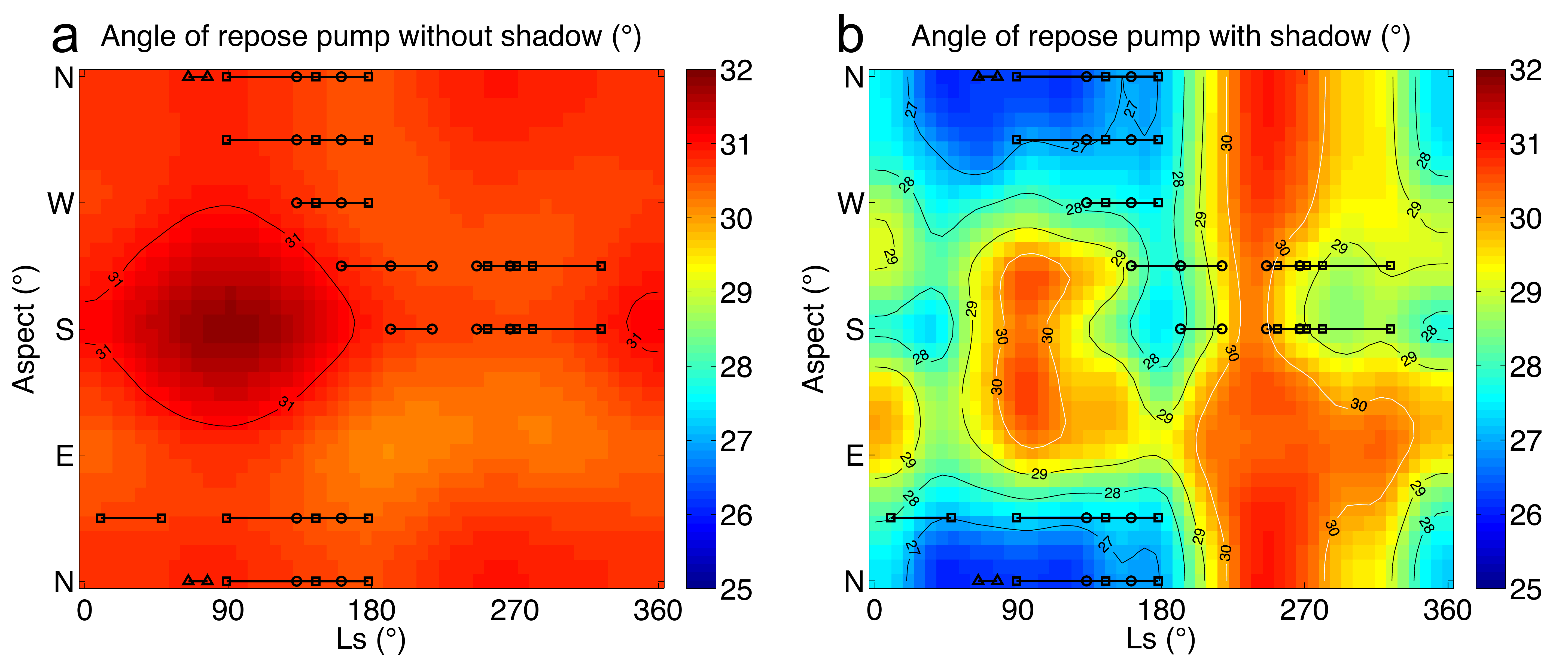}\caption{Angle of repose, modified by the Knudsen pump as a function of slope
aspect (expressed as the clockwise angle from the North) and season
(expressed in angle Ls). (a) case of a regular pump (b) case of an
enhanced pump. Black lines represents actual RSL activity as observed
by the HiRISE instrument in the Garni crater \citep{Chojnacki_Geologiccontextrecurring_JoGRP2016}.
Each activity is encompassed by two images: before and after. The
circles represent Martian year 31 (MY31) while the squares MY32. The
triangles represent the activity in the middle of the slope toward
North, without clear boulder at the origin of the RSL. \label{fig:Results_Repose_Aspect-Ls}}
\end{figure}

\paragraph*{Knudsen pump in the Martian soil}

Recently, some experimental works demonstrated that thermal creep
is able to lift grains at low pressure, only by irradiating the surface
\citep{Wurm_DustEruptionsby_PRL2006,Kelling_Dustejectionfrom_I2011}.
This photophoretic (or thermophoretic or thermal creep) effect on
the porous space of the soil (figure \ref{fig:Scheme-Dry-Flow} B)
has been proposed to be at the origin of dust lifting on Mars \citep{Wurm_Greenhouseandthermophoretic_GRL2008}.
A more precise experimental work at reduced gravity proved that this
process should occur on Mars in the form of a Knudsen pump \citep{Beule_martiansoilas_NP2013}.
The temperature field within few grains has been modeled and is enough
for photophoretic effect \citep{Kuepper_Photophoresispolydispersebasalt_JoAS2014}.
The precise near-surface temperature field in granular medium has
been estimated numerically \citep{Kocifaj_Dustejectionfrom_MNotRAS2010,Kocifaj_Radiativecoolingwithin_I2011},
taking into account the penetration depth of incoming light. On Mars,
rough estimates of the thermal balance demonstrate that the thermal
creep creates an active layer 100-200 microns thick for a particle
size of 67 microns \citep{Beule_insolationactivateddust_I2015,Kuepper_Thermalcreepassisted_JoGRP2015,Kuepper_Amplificationdustloading_I2016}.
This effect decreases by 10\% the wind speed threshold for saltation.
This effect has also been proposed to produce larger dust devils by
self shadowing of the dust devil itself. More generally, Knudsen pump
has been recognized to be very promising to understand geomorphological
features on Mars \citep{Schorghofer_PlanetaryscienceSubsurface_NP2014}
but until now, no specific applications have been proposed.

\paragraph*{Modified angle of repose}

We put forward that the photophoretic effect, triggered by the sun
radiation, modifies seasonally the angle of repose of the granular
material. Since the slopes of granular material are stable near 30\textdegree ,
even a minor change in the angle of repose due to the air flux could
significantly change the stability. Figure \ref{fig:Scheme-Dry-Flow}
(A) summarizes our model to trigger RSL flows, by a simple Knudsen
pump (regular pump) and by shadowing from a boulder (enhanced pump). 

Knowing the direct solar flux, the one scattered by the atmosphere,
and the thermal infrared flux from the Mars Climate Database \citep{Lewis_MCD_JGR1999},
we can estimate the slope effect \citep{Spiga_SolarIrradiance_GRL2008}
on the three incoming energy fluxes. By using these values as boundary
conditions on precise thermal surface balance model \citep{Kocifaj_Dustejectionfrom_MNotRAS2010,Kocifaj_Radiativecoolingwithin_I2011},
we aim at estimating both regular (without shadow) and enhanced (with
shadow) Knudsen pump effects on RSL sites \citep{Beule_insolationactivateddust_I2015,Kuepper_Thermalcreepassisted_JoGRP2015,Kuepper_Amplificationdustloading_I2016}.
Both regular and enhanced pumps are estimated in the most favorable
conditions, when the surface temperature is higher at maximum solar
elevation. Figure \ref{fig:Scheme-Dry-Flow} (C) shows an example
of a temperature profile for the regular pump (at local maximum solar
elevation). The maximum enhancement effect occurs when the thermal
gradient is maximum in the upper part, usually 1 to 5 minutes after
shadow over passed.

By adapting a dry granular flow model \citep{Pouliquen1999}, we estimate
the slope angle required to flow $\theta_{start}$ (limit angle of
repose to start the flow) from the pumping effect. In this model,
$\theta_{start}$ is in the range 23\textdegree -36\textdegree{} and
is dependent on the thickness of the flowing granular material. The
thickness is determined by the upper layer of the thermal profile.
Grain radius is set to 50 microns in agreement with the analysis of
the infrared observations \citep{Edwards_WaterContentRecurring_GRL2016}.
The observed slope is around 30\textdegree{} (see fig. \ref{fig:RSL-image}).
We choose to compute the model for the Garni crater, significantly
better documented than other places on Mars \citep{Chojnacki_Geologiccontextrecurring_JoGRP2016}
and offering a variety of slopes in a very localized place. We choose
thermal parameters in accordance with previous analysis \citep{Edwards_WaterContentRecurring_GRL2016}
(see sup. mat.). Since the observed slope is 30\textdegree , we will
consider here that an angle of repose strictly lower than 30\textdegree{}
will provoke the flow.

\paragraph*{Modelled flow activity on Mars}

We computed the modified angle of repose as a function of season and
slope orientation (aspect). The main results are summarized in fig.
\ref{fig:Results_Repose_Aspect-Ls}. The regular pump is never efficient
enough to trigger a flow. Conversely, if a boulder creates a shadow
near local maximum solar elevation, the enhanced pump is efficient
enough to trigger a flow by reducing the angle of repose down to 26.2\textdegree ,
significantly less than the observed slope. The comparison with actual
observations at Garni crater gives a very good agreement. The general
pattern of slopes towards South near Ls=270\textdegree{} and towards
North near Ls=90\textdegree{} is captured. Slopes towards East seem
always inactive, as predicted with our model. These slopes have a
lower thermal gradient due to less absorbed energy before the shadow
(at local maximum solar elevation) at 10AM in the East, instead of
noon in North/South and 14PM in the West. The flow activity in the
middle of the Northern slope at Ls=80\textdegree{} is in the local
maximum efficiency of the enhanced pump. We propose that self shadowing
of the granular material itself may be at the origin of this activity.

In summary, our new model is compatible with most observational facts:
(i) From HiRISE images, most of the sources of RSL are in rough terrains
with slope $\sim$30\textdegree . Highest slopes \textgreater{} 40\textdegree{}
remain unmodified. (ii) Most of the flow seems to originate from boulders.
(iii) The activity of RSL occurs at low slope angle (\textless{} 34\textdegree ),
lower than the usual angle of repose of wet/dry granular flow. We
propose that the required lubrication to flow is provided by the Knudsen
pump. (iv) The seasonality and slope facing activity is well reproduced
by our model. (v) We can observe a slight advantage in West facing
facets compared to East facing ones \citep{Chojnacki_Geologiccontextrecurring_JoGRP2016}.
We argue that Western slopes accumulate more heat leading them to
be at higher maximum temperature; moreover, even in shaded conditions
these slopes will have a higher temperature gradient. (vi) Our model
does not require water (nor CO$_{2}$ ice), which is difficult to
explain in the equatorial region of Mars. 

A new dry exotic flow, based on Knudsen pump, occurring only near
the equator of Mars is proposed to explain the cryptic RSL features.
It constitutes an alternative hypothesis without pure liquid water
nor brines. If confirmed, the absence of surface liquid would significantly
modify our understanding of the Martian environment and the current
habitability should to be reassessed. Future experimental studies
will be conducted in order to better quantify this process. As a perspective,
this exotic dry flow mechanism could also occur in other planetary
environments, in extremely rarefied gas such as those of Pluton, Triton
or even the Moon.

\section{Method}

\subsection{Irradiance on a sloped surface}

The direct solar irradiance $I_{0}$ reaching the top of the atmosphere
at the longitude, latitude, Ls and local time ($\delta,\phi,t$) is
computed using standard equations \citep{Schmidt_AlbedoSSPC_Icarus2009}.
The direct irradiance reaching the surface $D_{0}$ is deduced knowing
the local atmospheric opacity from the Mars Climate Database 5.2 \citep{Lewis_MCD_JGR1999,Forget_MarsGCM_JGR1999,Millour_MCD5.2_EPSC2015},
based on average TES dust optical retrieval \citep{Smith_TESAtmo_JGR2001}.
The total irradiance scattered by the atmosphere reaching the surface
$S_{0}$ is also estimated from the Mars Climate Database. 

For each facet, at a slope $\theta$, and azimuth of slope (or aspect)
$\xi$, the local incidence $\mu_{s}=\cos\alpha_{s}$ can be estimated
from the incidence $\mu=\cos\alpha$ on a flat facet:

\begin{equation}
\mu_{s}=\max\left(0,\mu_{0}\cos\theta+\sqrt{1-\mu_{0}^{2}}\sin\theta\cos\left(\xi-\xi_{0}\right)\right)\label{eq:incidence_angle}
\end{equation}

Using optical properties of the aerosols \citep{Ockert-Bell_Absorptionandscattering_JGR1997},
we can deduce the irradiance of the direct beam $D$, the beams reflected
by surrounding surfaces $R$ and the beams scattered by the atmosphere
$S$ \citep{Spiga_SolarIrradiance_GRL2008}. The total irradiance
reaching the facet is thus $I=D+R+S$.

The thermal infrared incidence flux $I_{T}$ is estimated from the
Mars Climate Database \citep{Lewis_MCD_JGR1999,Forget_MarsGCM_JGR1999,Millour_MCD5.2_EPSC2015}.

\subsection{Thermal model within the soil}

In order to determine the precise thermal profile in a granular material,
one has to solve the heat transfer equation, taking into account the
penetration of light in the near surface \citep{Kocifaj_Dustejectionfrom_MNotRAS2010,Kocifaj_Radiativecoolingwithin_I2011}.

\begin{equation}
k\frac{\partial^{2}T(t,h)}{\partial h^{2}}-\frac{\partial q_{r}(t,h)}{\partial h}=\rho C\frac{\partial T(t,h)}{\partial t}\label{eq:heat_transfer}
\end{equation}

where $T(t,h)$ is the temperature field depending on time and depth,
$k$ is the thermal conductivity, $\rho$ is the density, $C$ the
specific heat and $q_{r}$ is the radiative flux from direct attenuated
beam and diffused beam inside the soil. We estimate $q_{r}$ using
radiative transfer model \citep{Kocifaj_Dustejectionfrom_MNotRAS2010},
with optical properties of the grains \citep{Pollack_Opticalpropertiessome_I1973,Wolff_Wavelengthdependencedust_JGR2009}
and a grain radius of 50 microns. We solved eq. \ref{eq:heat_transfer}
numerically as previously described \citep{Kocifaj_Dustejectionfrom_MNotRAS2010,Kocifaj_Radiativecoolingwithin_I2011}. 

We simply adapt the model, initially designed for constant illumination,
to Martian daily illumination on a sloped facet by inserting $I$
and $I_{T}$ in the corresponding equations. 

\subsection{Knudsen pump model}

At low pressure, temperature gradients are very important to conduct
molecular flow in a tube. The Knudsen pump equation describes the
mass flow $\dot{M}$ through a capillary tube of length $L_{x}$,
derived from the linearized Boltzmann's equation \citep{Han_KnudenPump_Thesis2006,SONE_AnalysisPoiseuilleand_JoVSoJ1990}
: 

\begin{equation}
\dot{M}=P_{avg}A\sqrt{\frac{m}{2k_{B}T}}\left(\frac{L_{r}}{L_{x}}\frac{\Delta T}{T_{avg}}Q_{T}-\frac{L_{r}}{L_{x}}\frac{\Delta P}{P_{avg}}Q_{P}\right)\label{eq:Knudsen_flow}
\end{equation}

where $\Delta P$ is the pressure difference, $\Delta T$ is the temperature
difference, $P_{avg}$ is the average pressure, $T_{avg}$ is the
average temperature, $A$ is the capillary cross section area, $L_{r}$
is the radius of the capillary tube, $L_{x}$ is the length of the
capillary tube, $m$ the molecular mass of the gas, $k_{B}$ the Boltzmann's
constant, $Q_{T}$ is the thermally driven flow coefficient, $Q_{P}$
is the pressure driven return flow coefficient. Those coefficients
have been numerically evaluated \citep{SONE_AnalysisPoiseuilleand_JoVSoJ1990}. 

From thermal calculation, the temperature gradient of a granular medium
when the surface is irradiated can be considered as constant at the
near surface for a length $L_{x2}$ (upper domain 2) and then decreases
at depth for a length $L_{x1}$ (lower domain 1) \citep{Beule_insolationactivateddust_I2015}.
We can thus estimate the pressure increase at the limit between domain
1 and 2 \citep{Beule_insolationactivateddust_I2015}:

\begin{equation}
\Delta P=\frac{L_{x2}}{L_{x2}+L_{x1}}P_{avg}\frac{\Delta T_{1}}{T_{avg}}\frac{Q_{T}}{Q_{P}}\label{eq:pressure_regular}
\end{equation}

When the irradiation stops, surface temperature decreases, creating
a gradient in the upper domain 2, leading finally to an amplification
of the Knudsen pump \citep{Kuepper_Amplificationdustloading_I2016}.
The enhancement factor is \citep{Kuepper_Amplificationdustloading_I2016}: 

\begin{equation}
E=\frac{\Delta P_{enhanced}}{\Delta P}=1+\frac{\Delta T_{2}}{\Delta T_{1}}\frac{L_{x1}}{L_{x2}}\label{eq:enhancement_factor_pressure}
\end{equation}

\subsection{Flow model}

A fundamental aspect of granular flow is that there is an hysteresis
in the slope stability: a metastable state exists between the static
an the flowing states, where a flow can be triggered by a finite disturbance
\citep{Mangeney_Numericalmodelingself_JoGR2007}. In this article,
the physical modeling of the destabilization of the material and the
granular flow is based on an empirical formulation of a Coulomb-type
friction law \citep{Pouliquen1999,POULIQUEN2002}, built on two functions
$h_{stop}\left(\theta\right)$ and $h_{start}\left(\theta\right)$
that are determined experimentally. $h_{stop}\left(\theta\right)$
is the thickness of granular media necessary to observe a steady uniform
flow on an inclined plane at the inclination $\theta$, and $h_{start}\left(\theta\right)$
is the minimum thickness of a granular layer necessary to generate
a flow on the same plane. 

To take into account the addition of an uplifting force due to the
Knudsen pump effect, we chose to consider a rotation of the coordinate
system, as if the lifting force that is perpendicular to the surface
acted as a change in the gravity field. This approach is justified
because the forces that apply to the flowing layer are depth averaged
in the physical model considered \citep{Mangeney_Numericalmodelingself_JoGR2007}.

For a given slope angle $\theta$ (considered here to be 30\textdegree ,
as determined using the DTM), we determine if an uplifting force $F_{lift}$,
caused by the Knudsen pump, can destabilize a granular layer of thickness
$h=L_{x2}$. Considering that the granular material in the slopes
has friction properties similar to sand, \citep{POULIQUEN2002,Felix_DryGranularFlow_EPSL2004,Mangeney_Numericalmodelingself_JoGR2007}
we estimate the critical angle $\theta_{start}$:
\begin{equation}
\tan\theta_{start}=\frac{L\tan\theta_{4}+h\tan\theta_{3}}{L+h}
\end{equation}
with $\theta_{3}$, $\theta_{4}$ and $L$ determined experimentally.
Taking into account the lifting force $F_{lift}$ and the gravity
force $F_{grav}$, we estimate the equivalent slope angle $\theta_{eq}$
using: 
\begin{equation}
\tan\left(\theta_{eq}-\theta\right)=\frac{F_{lift}\sin\theta}{F_{grav}-F_{lift}\cos\theta}\label{eq:thetaeq}
\end{equation}
The layer is destabilized by the Knudsen pump effect if $\theta_{eq}>\theta_{start}$. 

\subsection{Angle of repose}

We compute the angle of repose as a function of (i) season from Ls=0\textdegree{}
to 330\textdegree{} each 30\textdegree , which is the resolution of
the Mars Climate Database, and aspect from 0\textdegree{} to 315\textdegree{}
each 45\textdegree . The angle of repose represents the angle to start
a flow. In the case of the enhanced pump, it represents the angle
of repose of a soil when shadow passed over, near the local maximum
solar elevation, for instance due to the presence of a boulder.

\subsection{Data availability}

The authors declare that the data supporting the findings of this study are available within the article and its supplementary information files.

\subsection{Code availability}

The code is available from the corresponding author upon request.


\subsection*{Acknowledgement}

F.S. is the corresponding author. We acknowledge support from the
``Institut National des Sciences de l'Univers'' (INSU), the \char`\"{}Centre
National de la Recherche Scientifique\char`\"{} (CNRS) and \char`\"{}Centre
National d'Etude Spatiale\char`\"{} (CNES) through the \char`\"{}Programme
National de Plan{\'e}tologie\char`\"{}, HRSC/MEX, OMEGA/MEX and PFS/MEX
programs. Computational work was supported by the Slovak Research
and Development Agency under the contract No. APVV-14-0017. This work
is supported by the Center for Data Science, funded by the IDEX Paris-Saclay,
ANR-11-IDEX-0003-02. We thank Sylvain Piqueux and an anonymous reviewer
for their fruitful remarks.

\subsection*{Corresponding author}

Fr{\'e}d{\'e}ric Schmidt (frederic.schmidt{[}@{]}u-psud.fr)

\subsection*{Author contributions}

F.S has led the project. F.A. was in charge of the angle of repose
theory. F.C. has interpreted the RSL geomorphology. M.K., A.G.M. and
F.S. adapted the thermal profile calculation. All co-authors contributed
to the analysis of the results and the redaction of the article. 

\subsection*{Competing financial interests}

The authors declare no competing financial interests.

\end{document}